\documentclass[a4paper,11pt]{article}

\usepackage{amsmath}
\usepackage{amssymb}
\usepackage{graphicx}
\usepackage{subfig}
\usepackage{psfrag}
\usepackage{fullpage}
\usepackage{color}
\usepackage{pgf}
\usepackage[switch]{lineno}
\usepackage{authblk}
\usepackage{setspace}

\DeclareMathOperator{\csch}{csch}

\newcommand{\bs}[1]{\boldsymbol{#1}}

\author[1]{William D.\ Rowley}
\author[1]{William J.\ Parnell}
\author[2]{I.\ David Abrahams}
\author[3]{S.\ Ruth Voisey}
\author[3]{John Lamb}
\author[3]{Nicolas Etaix}

\affil[1]{\small{School of Mathematics, University of Manchester, Oxford Road, Manchester, M13 9PL, UK}}
\affil[2]{\small{Isaac Newton Institute, University of Cambridge, 20 Clarkson Road, Cambridge CB3 0EH, UK}}
\affil[3]{\small{Dyson Technology Limited, Tetbury Hill, Malmesbury SN16 0RP, UK}}

\title{\textbf{Deepening subwavelength acoustic resonance via metamaterials with universal broadband elliptical microstructure}}

\date{}

\definecolor{dkgreen}{rgb}{0.05,0.25,0.1}
\definecolor{gray}{rgb}{0.5,0.5,0.5}
\definecolor{mauve}{rgb}{0.58,0,0.82}
\definecolor{lightgray}{rgb}{0.99,0.99,0.99}

\begin{document}

 \maketitle

    \begin{abstract}
Slow sound is a frequently exploited phenomenon that metamaterials can induce in order to permit wave energy compression, redirection, imaging, sound absorption and other special functionalities. Generally however such slow sound structures have a poor impedance match to air, particularly at low frequencies, and consequently exhibit strong transmission only in narrow frequency ranges. This therefore strongly restricts their application in wave manipulation devices. In this work we design a slow sound medium that halves the effective speed of sound in air over a wide range of low frequencies, whilst simultaneously maintaining a near impedance match to air. This is achieved with a rectangular array of cylinders of elliptical cross section, a microstructure that is motivated by combining transformation acoustics with homogenization. Microstructural parameters are optimised in order to provide the required anisotropic material properties as well as near impedance matching. We then employ this microstructure in order to halve the size of a quarter-wavelength resonator (QWR), or equivalently to halve the resonant frequency of a QWR of a given size. This provides significant space savings in the context of low-frequency tonal noise attenuation in confined environments where the absorbing material is adjacent to the region in which sound propagates, such as in a duct. We term the elliptical microstructure `universal' since it may be employed in a number of diverse applications.
    \end{abstract}

\newpage

The ability to shape, redirect and manipulate sound has been of interest for many decades and the recent emergence of new composite materials and metamaterials has driven this area forward with a multitude of exciting results including cloaking, negative refraction and lensing \cite{cummer2016controlling, cummer2007one, torrent2008acoustic, zhang2009focusing, christensen2012anisotropic, kaina2015negative, groby2015use}. The application of slow sound via space-coiling, labyrinthe type structures \cite{liang2012extreme, liang2013space, xie2013measurement, frenzel2013three} and helical devices \cite{zhu2016implementation} is of significant interest since, as with slow light, the concept has great promise in a number of scenarios. In many applications however, when the surrounding material is air, one requires full, broadband transmission through slow sound devices in order to enable complete manipulation of the sound field via a metamaterial. This is not the case in general for previously developed materials \cite{liang2013space, zhu2016implementation, climente2010sound}.

In this Letter we use the method of transformation acoustics \cite{cummerChapter, norris2008acoustic} to design a medium in which space is apparently stretched or alternatively sound is effectively slowed, whilst also ensuring that the medium is almost-impedance matched to air. We employ a recently developed homogenization method in order to realise this material via a microstructure consisting of rigid elliptical cylinders arranged in a rectangular array. The explicit form of anisotropic density provided via the homogenization scheme means that microstructures can be optimised to best effect. This microstructure may be contrasted with perforated plates, which have been the focus of the majority of previous studies, and have been employed to realise cloaks \cite{zhang2011broadband}, ground cloaks \cite{popa2011experimental, zigoneanu2014three}, lensing \cite{christensen2012anisotropic} and right angle bend transformations \cite{lu2017design}. The newly proposed elliptical microstructure achieves a near impedance match to air at a broad range of low frequencies and furthermore induces an effective sound speed in the heterogeneous medium that can be reduced to one half that of air. These achievements and generalizations of the microstructure  can be exploited in a number of important applications and therefore we term the microstructure `universal'. Here we focus on the application of low-frequency resonance and exploit our findings to halve the size of the well known quarter-wavelength resonator (QWR) \cite{kinsler1999fundamentals}. A QWR is a side branch of an acoustic duct used to attenuate tonal noise of wavelength approximately four times the length of the side branch; this attenuation is achieved through destructive interference.   Theoretical predictions are validated experimentally; the resonance peak is shifted close to that predicted by the theory, associated with the reduction in sound speed in the resonator. The amplitude of the resonance is reduced somewhat due to inherent viscous and thermal effects in the heterogeneous structure, but the ability to significantly reduce the resonant frequency (or equivalently reduce the size of the resonator) is undoubtedly of significant practical importance.

More broadly speaking, along with Helmholtz resonators and Herschel-Quincke tubes, QWRs are frequently used in applications where low-frequency tonal noise needs to be attenuated in a confined environment such as a duct. Manipulating acoustic waves in such contexts where sound propagates orthogonal to the walls of the confined region is traditionally very difficult. Although high-frequency sound can easily be attenuated via standard sound-absorbing foams, low-frequency sound is more difficult to attenuate since it requires prohibitively large amounts of such materials. The standard approach for low-frequency sound attenuation in ducts is therefore to employ resonators and side-branches \cite{lee2009acoustic, wang2012wave} and although such side branch resonators can be useful, their size frequently prohibits practical use, although some work has been done using standard resonators in sequence \cite{lee2009acoustic, wang2012wave, fang2006ultrasonic}. Generally there is a lack of work in the area of novel metamaterial approaches applied to sound attenuation in confined spaces; furthermore the resonators employed would often impede flow e.g.\ \cite{lee2010composite}. This is in contrast to sound absorption in free-space where sound is incident onto a surface: the concept of metasurfaces has received significant attention, with many works describing structures capable of perfect sound absorption at low-frequency e.g.\ \cite{mei2012dark, jimenez2016ultra}. Unfortunately, this concept cannot be of use in the same manner in the context of propagation in confined regions, since the metasurface would either sit across the duct, orthogonal to the direction of the sound field, perform well but obstruct flow, or sit parallel to the direction of sound, and be ineffective, whilst permitting flow. The size restriction associated with side-branch resonators means therefore that the present finding could be of broad significance in low-frequency noise control.


As described above, a range of necessary material properties derived through transformation acoustics have been realised in a variety of metamaterial applications and have been experimentally verified to good effect. Here, an effective elongation of space is provided by employing one of the simplest mappings in the theory of transformation acoustics. The transformation from the virtual primed coordinate system $(x',y',z')$ onto the physical system $(x,y,z)$ is $x=x', y=\alpha y', z=\beta z'$. We are interested in sound propagation in the two-dimensional $xy$-plane, with $\alpha<1$, which creates an apparent elongation of space via a reduction in the sound speed (see Fig.\ 1(a)). Microstructure will subsequently be chosen in order to realise this scenario. The parameter $\beta$ is employed here to optimise impedance matching to air whilst still permitting realistic microstructures.

With $\rho'$ and $K'$ denoting the acoustic density and bulk modulus of air, transformation acoustics dictates the required material properties associated with the mapping. These required properties take the form of an anisotropic density with components $\rho_x^{\star}$ and $\rho_y^{\star}$ and a scalar bulk modulus $K^{\star}$ (see Supplemental Material). Note that we do not concern ourselves with matching the required density in the $z$-direction, and so disregard any stretch or confinement in the $z$-direction; this will lead to an impedance mismatch that can be quantified through the value of $\beta$. With $\beta=1$, $\rho_y^{\star}$ is required to be less than $\rho'$. This would provide a perfect impedance match but naturally raises fabrication issues. Employing $\beta>1$ allows an increase in $\rho_y^{\star}$, above $\rho'$, and therefore ensures more straightforward fabrication but with a slight impedance mismatch. This approach was suggested by Popa et al.\ \cite{popa2011experimental} in the context of the two-dimensional acoustic ground cloak.

\begin{figure}[htb]
\centering
\includegraphics[width=.5\textwidth]{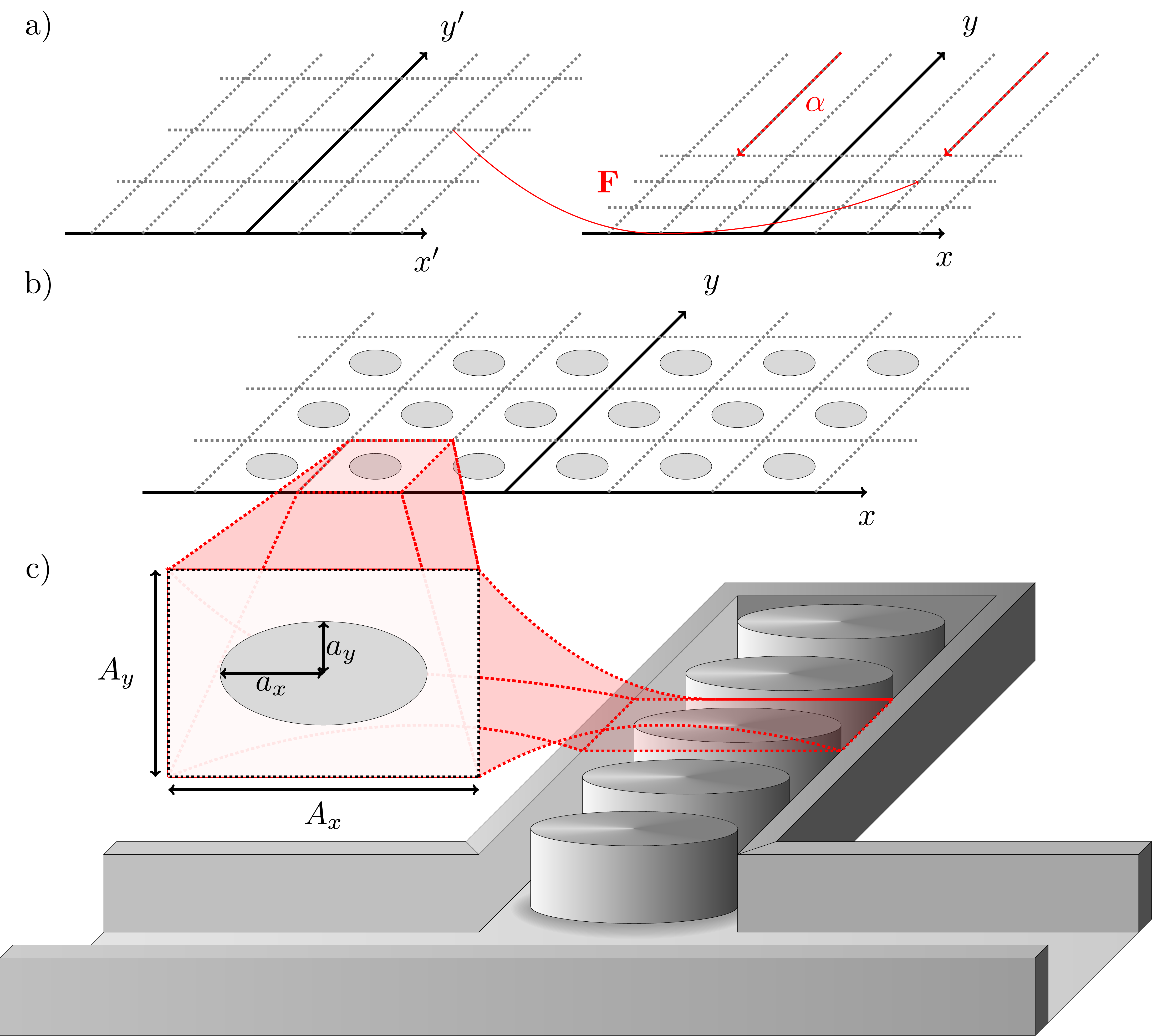}
\caption{a) Visualisation of the transformation on the $xy$-plane from the uniform virtual space onto physical space used to obtain the required material properties. b) The physical `stretched' space is realised by the employment of rigid elliptical cylinders placed on a rectanglar lattice as is illustrated here together with the unit cell employed. c) Illustrating that the microstructure realised in (b) can be employed in the quarter wavelength resonator to halve its length, whilst ensuring that it operates at the same frequency.}\label{fig:schematic}
\end{figure}

In order to realise the required properties we employ an array of rigid cylinders with elliptical cross-section arranged on a rectangular lattice. The ellipse has semi-axes $a_x$ and $a_y$ and the array has lattice spacings of $A_x$ and $A_y$  along the $x$ and $y$ directions respectively as shown in Fig.\ \ref{fig:schematic}(b). Parameters need to be chosen such that the effective material properties of this inhomogeneous medium, which we shall denote as $\rho_x^{\mathrm{eff}}, \rho_y^{\mathrm{eff}}$ and $K^{\mathrm{eff}}$, match those of the required material arising from transformation acoustics, i.e.\ $\rho_x^{\star}=\rho_x^{\textnormal{eff}}, \rho_y^{\star}=\rho_y^{\textnormal{eff}}$ and $K^{\star}=K^{\textnormal{eff}}$, but in general this is not possible for the application considered here. We therefore follow a similar approach to Popa et al \cite{popa2011experimental} and minimise the difference between the material properties of a structure we can readily fabricate and those of the required material with $\beta$ as an additional parameter associated with impedance mismatch. The problem therefore possesses five parameters, two belonging to the transformation, $\alpha$ and $\beta$, and three belonging to the elliptical-microstructure, $a_x$, $a_y$, the semi-axes, and $A=A_x/A_y$ the relative spacing. The lengthscale $A_y$ is the size of the microstructure, which is required to be much smaller than the propagating wavelengths of interest for homogenization theory to be valid.

The effective properties of the array are determined by the integral equation method of homogenization \cite{parnell2008,joyce2017integral}, the benefit of which is an explicit form for the effective properties in terms of the parameters of the problem. For fast optimisation we use the simplest version of the method to determine the domain in which the optimal properties will reside. We then use Comsol Multiphysics to further explore this local region of parameter space identified by the homogenization method, including viscous and thermal losses to provide an optimal parameter choice (see Supplemental Material).

Using this approach, we minimise the difference between the effective properties of a medium with a given ellipse of fixed semi-axes, and those required by a general transformation over the transformation parameters and relative lattice spacing. We find an optimal microstructure yielding $\alpha= 0.5$, with $a_x=0.47 A_x, a_y=0.05 A_y, A_x=2.67A_y$. This gives effective densities of approximately 1.1 and 5.0 times the density of air in the $x$ and $y$ directions respectively; the bulk modulus is approximately 1.1 times that of air also. Hence in the $y$ direction the speed of sound (given by $\sqrt{K^\mathrm{eff}\rho_y^\mathrm{eff}}$) is reduced to approximately half that of air, while in the $x$ direction the speed of sound is approximately matched to that of air. It remains to choose a value for $A_y$, which should be significantly smaller than the incident wavelength in order for the homogenization theory to be valid.  However, if we take $A_y$ too small we will greatly increase the surface area of the embedded microstructure and promote viscous and thermal losses within our effective material, which may not be ideal (see Supplemental Material). The selection of this final constant is application dependent. With the parameters as chosen above, we term this a `near-miss' microstructure, referring to its ability to closely match the required material properties from transformation acoustics.

Having optimised the microstructure to reduce sound speed inside the inhomogeneous medium to half that of air, we now employ this medium in the application to a QWR of length $L$, as illustrated in Fig.\ \ref{fig:schematic}(c). Without microstructure the QWR will naturally attenuate waves of wavelength approximately $\lambda=4L$ corresponding to a frequency of attenuation $f=c/\lambda$ where $\lambda$ is the wavelength of the incident wave in air and $c$ the speed of sound in air. Here however, by employing the slow sound medium with near impedance matching, for the same length $L$ we achieve $f=c/2\lambda$, hence deepening the resonant frequency significantly and effectively creating an \textit{eighth wavelength resonator}. An alternative way of viewing this is to say that with the microstructured resonator one can attenuate waves of the same frequency with a resonator that is half the length of a standard QWR, thus creating a space-saving resonator device.

\begin{figure}[h!!]
	\centerline{\includegraphics[width=.7\textwidth]{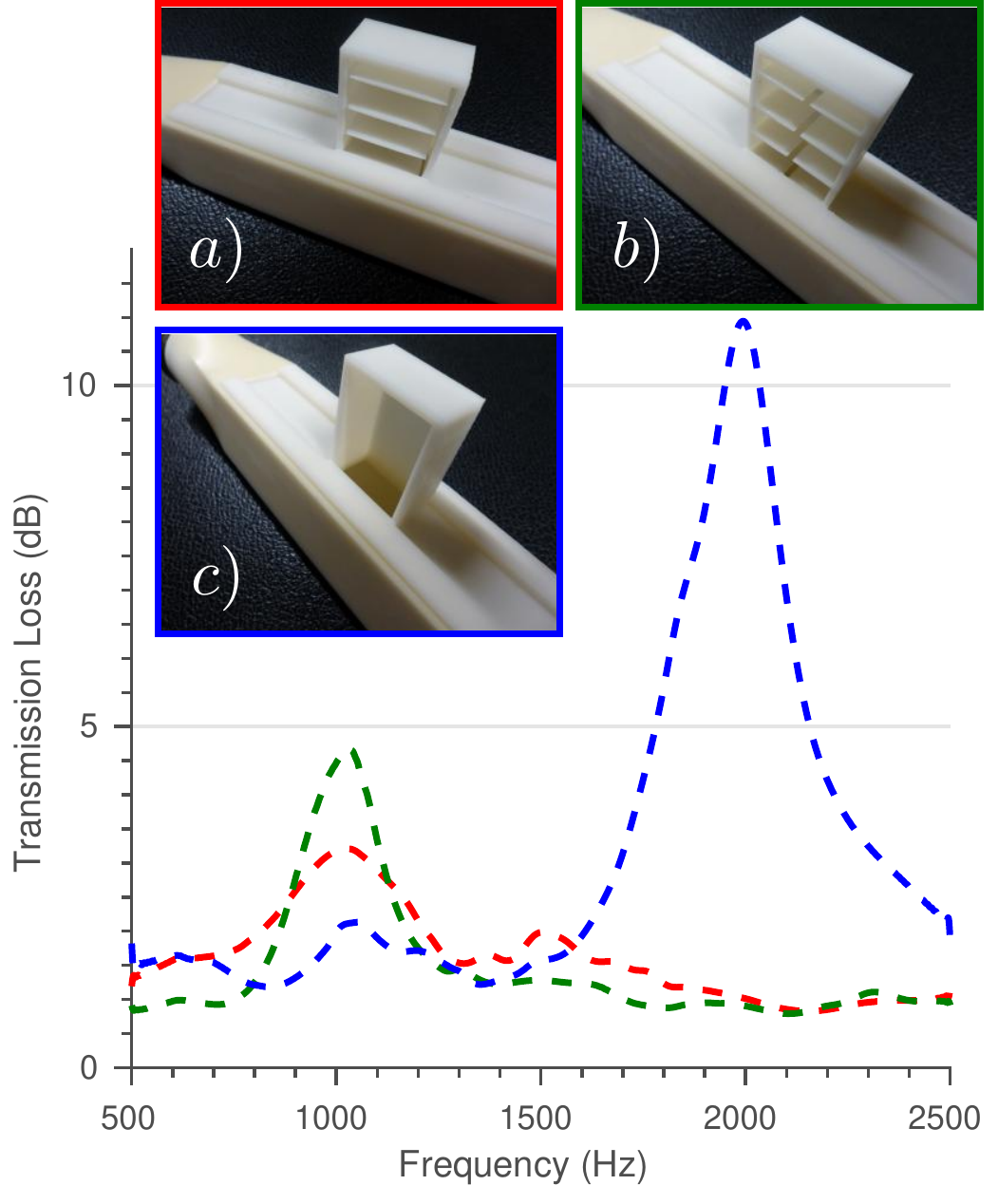}}
	\caption{Experimental results illustrating the transmission loss across three resonant side branches of equal length. Branch (a), corresponding to the red curve, has a microstructure of one ellipse wide with the ellipse centred on the branch centre. Each unit cell is 26.7mm wide and 10mm tall; the ellipse has a major semi-axis of 12.5 mm and a minor semi-axis of only 0.5mm. There are four unit cells within the side branch making the side branch 40mm deep with an opening of 26.7mm. Branch (b), corresponding to the green curve, once again has a microstructure one ellipse wide, however now the ellipses are centred on the branch walls. All dimensions are as in the previous case, the unit cell is simply shifted. Branch (c), corresponding to the blue curve, is a standard air filled resonator for comparison. The data shown here has had noise removed using a fast Fourier transform. The halving of the resonant frequency in the case of a microstructured resonator is clearly visible.}\label{fig:results}
\end{figure}

Employing a 3D printer, we fabricated a microstructured resonator of total length $L=40mm$, with cell height $A_y=10mm$ and with the other parameters as detailed above, for experimental testing in a standard two port experiment \cite{Experimental1,Experimental2}. The associated transmission loss, defined in the standard manner as $TL = 20 \log_{10}\left(\left|p_{in}/p_{out}\right|\right)$ (where $p_{in}$ is the total pressure field prior to the side branch and $p_{out}$ is the total pressure field after the side branch) is plotted as a function of frequency in Fig.\ \ref{fig:results}. The data shown has been post-processed using a Fourier transform to remove noise for clarity. The blue curve in the figure shows the transmission loss due to an ordinary air filled side branch, a QWR; we see a large amount of transmission loss at the predicted frequency, that corresponding to a wavelength of approximately four times the QWR's length. For the case of a single ellipse across the width of a side-branch of the same length (red curve) we clearly see the decrease in resonant frequency by a factor of a half as predicted, although as may be observed, the magnitude of the loss is decreased somewhat. This reduction is almost certainly associated with viscous and thermal losses in boundary layers in the narrow air gaps of the microstructure between the walls of the side branch and the ellipses. This loss mechanism results in less energy returning to the resonator neck in order to destructively interfere with the incoming wave at the resonant frequency. With this in mind a further resonator was printed with the microstructure shifted by a half cell width in order to have one wider air gap along the centre of the resonator rather than two narrower air gaps at the resonator walls. This offset microstructure does indeed give significant gain in transmission loss (green curve).

In conclusion, we have shown how to design and construct a slow sound material with near impedance match to air. Transformation acoustics is employed to derive the required material properties due to our prescribed coordinate stretch. Following this, the integral equation method of homogenization was pivotal in yielding approximate analytical formulae for the material properties of an array of elliptical cylinders on a rectangular lattice. These formulae allowed for fast and accurate optimisation of the microstructure in order to best fit the required slow sound properties. This approximation was then used as a starting point for numerical simulations, vastly narrowing down the search space in which to find an optimal configuration and including the effects of viscous and thermal losses. The developed slow sound material was used inside a quarter wavelength resonator in order to halve its resonant frequency, or alternatively to halve the length of a resonator required to attenuate a given tone. This specific application has the potential for significant industrial impact, specifically improved sound quality with a reduced form factor (patent pending \cite{ourpatent}); while the theoretical ideas and associated universal broadband microstructure has widespread potential for exploitation in a breadth of acoustic devices in the future.


{
\footnotesize
\bibliography{Rowley_et_al_for_submission_arxiv}
\bibliographystyle{unsrt}
}

\noindent
\textbf{Acknowledgements} WDR acknowledges the EPSRC and Dyson for funding via a KTN Industrial CASE PhD Studentship, WJP is grateful to the EPSRC for his fellowship grant EP/L018039/1, IDA contributed whilst in receipt of a Royal Society Wolfson Research Merit Award, and latterly under EPSRC grant EP/K033208/I.

\noindent
\textbf{Author Contributions}
WDR carried out initial studies into transformation acoustics. WJP and WDR worked on the associated homogenisation tools. WDR carried out analytical optimisation between required and effective properties and further numerical optimisation. WJP and IDA supervised the project. The QWR application was conceived by JL. SRV acted as the industrial supervisor to the project and facilitated all interactions between the University of Manchester and Dyson. Initial experimental testing was carried out by WDR and SRV. Optimisation in Comsol Multiphysics and final experiments were carried out by NE. All authors have discussed and contributed towards the final writing of this manuscript.

\newpage
\subsection*{Supplemental material}

\subsubsection*{Transformation acoustics}

The simple transformation used in order to derive the necessary material properties for a `slow sound' or alternatively an `effectively elongated', material is given by
\begin{equation}
x = x',\quad
y = \alpha y', \quad
z = \beta z',
\end{equation}
where we use the convention that primed coordinates belong to the virtual space and unprimed to the physical. The resulting Jacobian of transformation, $\boldsymbol{F}$, and associated determinant, $J$ are
\begin{equation}
\boldsymbol{F} = \left(\begin{array}{ccc}
\partial x/\partial x' & \partial x/\partial y' & \partial x/\partial z' \\
\partial y/\partial x' & \partial y/\partial y' & \partial y/\partial z' \\
\partial z/\partial x' & \partial z/\partial y' & \partial z/\partial z'
\end{array}\right)= \left(\begin{array}{ccc}
1 & 0 & 0 \\
0 & \alpha & 0 \\
0 & 0 & \beta
\end{array}\right),
\end{equation}
\begin{equation}
 \quad J = \det\boldsymbol{F} = \alpha \beta.
 \end{equation}
Transformation acoustics then gives the required material properties according to the formulae,
\begin{equation}
\left(\bs{\rho^\star}\right)^{-1} = \frac{1}{J \rho'}\boldsymbol{F F}^T, \quad K^\star = JK',
\end{equation}
in our case
\begin{equation}
\left(\bs{\rho^\star}\right)^{-1} = \left(\begin{array}{ccc}
1/\alpha\beta & 0 & 0 \\
0 & \alpha/\beta & 0 \\
0 & 0 & \beta/\alpha
\end{array}\right)\frac{1}{\rho'}, \quad K^\star = \alpha \beta K'.
\end{equation}
We define the various directional components of the required density matrix, and restrict our study to the first two dimensions, so that the properties to be met in order to achieve the slow sound material are
\begin{align}
\rho_x^{\star} &= \alpha\beta\rho', \label{eq:reqrhoy}\\
\rho_y^{\star} &= \frac{\beta}{\alpha}\rho', \label{eq:reqrhox}\\
K^{\star} &= \alpha\beta K'.\label{eq:reqBM}
\end{align}
We will attempt to realise these two-dimensional properties by considering an inhomogeneous medium and corresponding expressions for its effective properties derived via homogenization theory. We consider an array of cylindrical columns in order to fabricate the medium in three dimensions. This leads to an impedance mismatch since the explicit $z$ density is not realised. However this `error' is quantifiable through the value of $\beta$, and we anticipate that provided $\beta$ remains relatively close to unity, the device should perform as required.

\subsubsection*{Homogenization}

Recently a new method of homogenization called the `Integral Equation method' (IEM) was devised for the determination of the effective (acoustic) properties of a medium consisting of cylindrical fibres/rods on a periodic lattice \cite{joyce2017integral} based on the original work in \cite{parnell2008}. When the rods are rigid, have elliptical cross-section with semi-axes $a_x$ and $a_y$ along the $x$ and $y$ directions, and are arranged on a rectangular lattice of period $A$ in the $x$ direction and unity in the $y$ direction, the effective densities are predicted at leading order in terms of the volume filling fraction $\phi$ and given by
\begin{align}
\rho_x^{\textnormal{eff}} &= \left(\frac{1-B + \phi(1+S_1)}{1+B - \phi(1-S_1)}\right)\rho', \label{eq:effrhox}\\
\rho_y^{\textnormal{eff}} &= \left(\frac{1-B + \phi(1+S_2)}{1+B - \phi(1-S_2)}\right)\rho',\label{eq:effrhoy}
\end{align}
where $B=(1-\epsilon)/(1+\epsilon), \epsilon=a_x/a_y$ and $S_1$ and $S_2$ are lattice-sums, details of which are provided below. By a leading order approximation in terms of filling fraction, we mean that all terms of order $\phi^2$ and higher are removed from the denominator and numerator in expressions \eqref{eq:effrhox}-\eqref{eq:effrhoy}; this is not a dilute approximation, see \cite{parnell2008,joyce2017integral}. Further the well-known result
\begin{equation}
K^{\textnormal{eff}}=K'/(1-\phi) \label{eq:effBM}
\end{equation}
is also derived. The explicit form of effective densities allows for extremely rapid optimisation of material parameters, in contrast to computational homogenization schemes, where trial-and-error often plays a dominant role in the realisation process.

\begin{figure}[htb]
\centerline{\includegraphics[width=.5\textwidth]{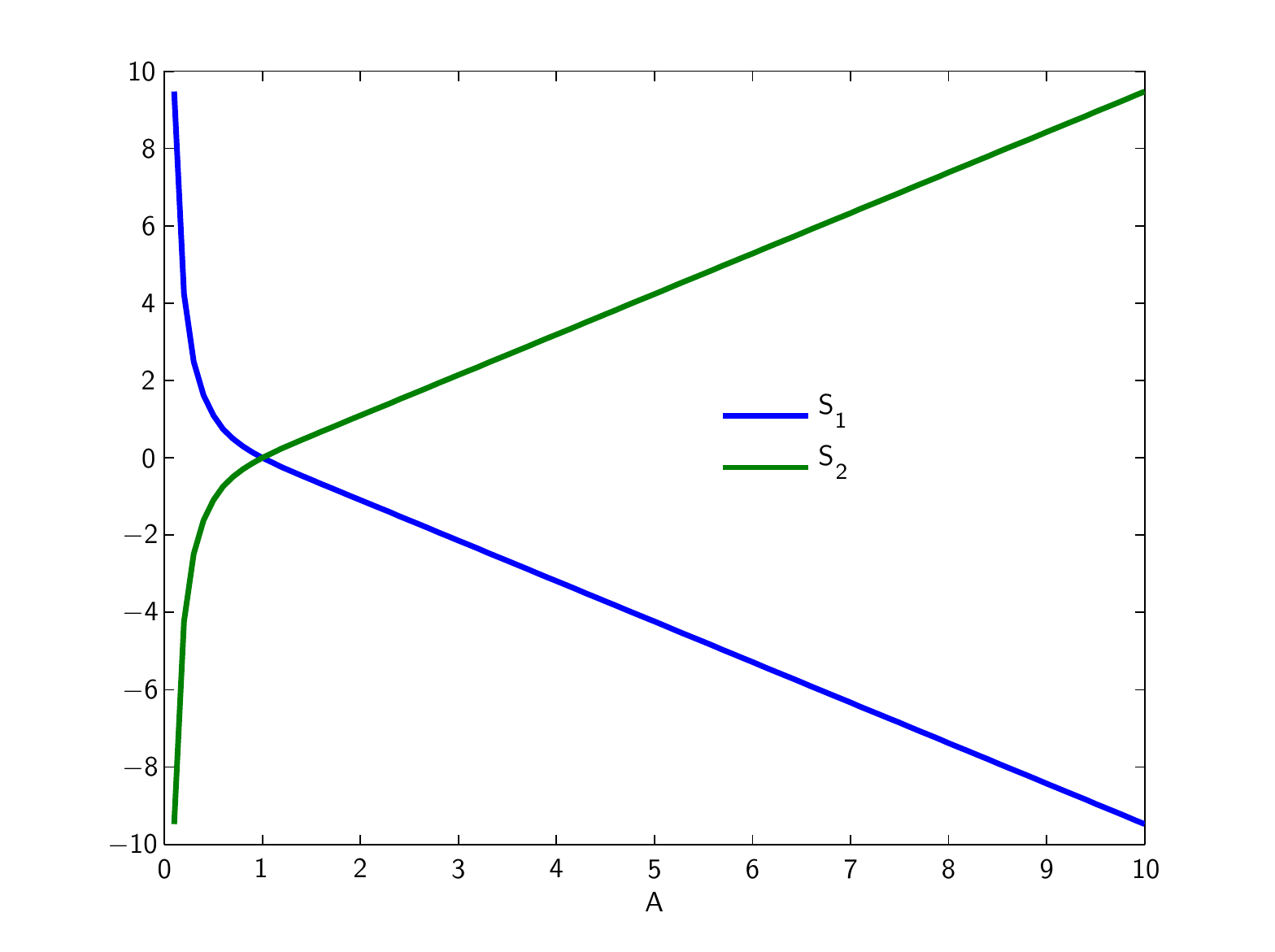}}
	\caption{Plot of the lattice sums $S_1$ and $S_2$ as a function of the lattice width $A$.}\label{fig:s1s2}
\end{figure}

The values of $S_1$ and $S_2$ can be calculated according to the formulae
\begin{align}
S_1 &= 1- \frac{\pi A}{3}\left(1- 6 \sum_{n=1}^{\infty} \csch^2(\pi n A)\right),\\
S_2 &= 1- \frac{\pi}{3A}\left(1- 6 \sum_{n=1}^{\infty} \csch^2\left(\frac{\pi n}{ A}\right)\right),\label{eq:secondLatticeSum}
\end{align}
where $A=A_x/A_y$ and $S_1$ and $S_2$ are illustrated in Fig.\ \ref{fig:s1s2}. Note that for $A=1$ both sums are equal to zero. This is the case of a square lattice in which the rectangular lattice sum disappears and the analysis is somewhat simplified.

\subsubsection*{Near-miss microstructures}

Due to the use of the IEM of homogenization we have approximate explicit forms for the effective material properties of the rectangular array of elliptical rods in terms of the physical geometry \eqref{eq:effrhox} - \eqref{eq:secondLatticeSum}. From transformation acoustics we also have the functional form of the required properties in terms of the transformation parameters \eqref{eq:reqrhox}-\eqref{eq:reqBM}. We may then consider the difference between these quantities in the sense of the $L_2$ norm and attempt to minimise this error. Provided that this error is sufficiently small, the elliptical microstructure should give a physical effect that is similar to that predicted by transformation acoustics with the determined values of $\alpha$ and $\beta$. To this end we consider ellipses with semi-axes between 0.01 and 0.49 (in increments of 0.01 due to fabrication tolerances) of the cell side length and minimise the quantity
\begin{equation}
\left|\dfrac{\boldsymbol{M}^\star\left(a_x^i,a_y^j,A\right) - \boldsymbol{M}^\mathrm{eff}\left(\alpha,\beta\right)}{\boldsymbol{M}^\star\left(a_x^i,a_y^j,A\right)}\right|
\end{equation}
over the parameters $A$, $\alpha$ and $\beta$. Where $\boldsymbol{M}$ is a vector of material properties, $\boldsymbol{M}=(\rho_x,\rho_y,K)$ while $a_x^i$ and $a_y^j$ are the semi axis of the $ij^\mathrm{th}$ ellipse considered, given by
\begin{align}
a_x^i = A i,  \label{axi} \\
a_y^j = j.     \label{ayj}
\end{align}
and $i$ and $j$ run between $0$ and $0.5$ in increments of $0.01$.

\begin{figure}
\centerline{\includegraphics[width=.5\textwidth]{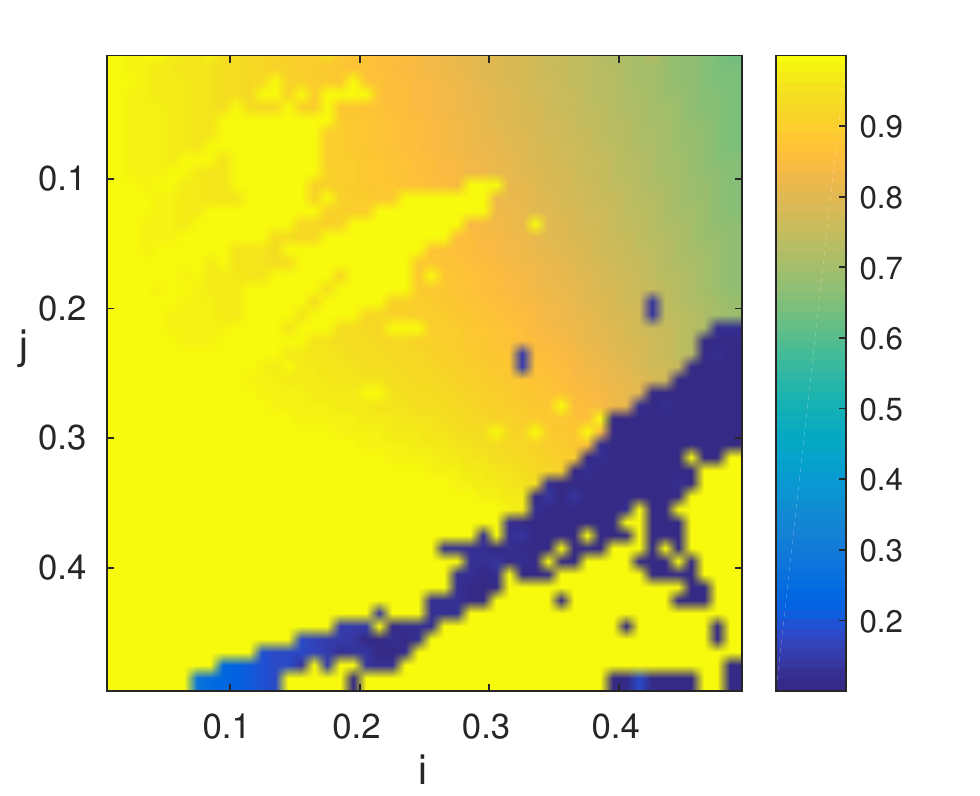}}
\caption{Illustrating the various $\alpha$ values that can be achieved by employing an elliptical microstructure.}\label{fig:alphas}
\end{figure}

\begin{figure}
\centerline{\includegraphics[width=.5\textwidth]{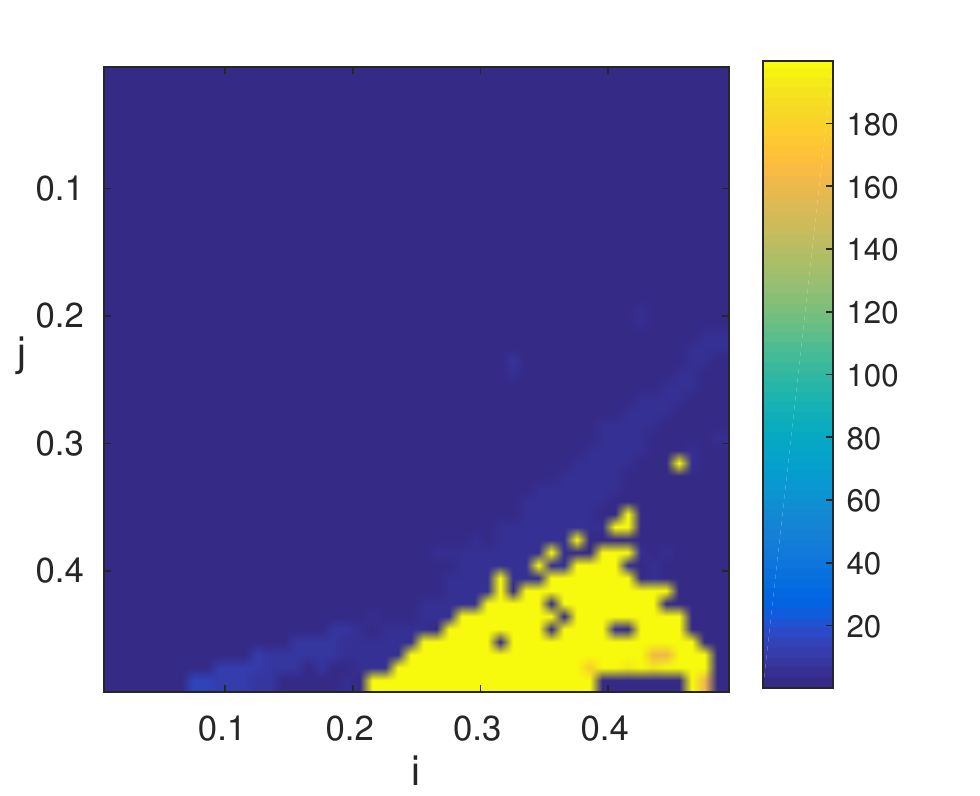}}
\caption{Illustrating the necessary cell width $A$, required for a given ellipse to achieve the $\alpha$ scaling in figure \ref{fig:alphas}.}\label{fig:Avalue}
\end{figure}

\begin{figure}
\centerline{\includegraphics[width=.5\textwidth]{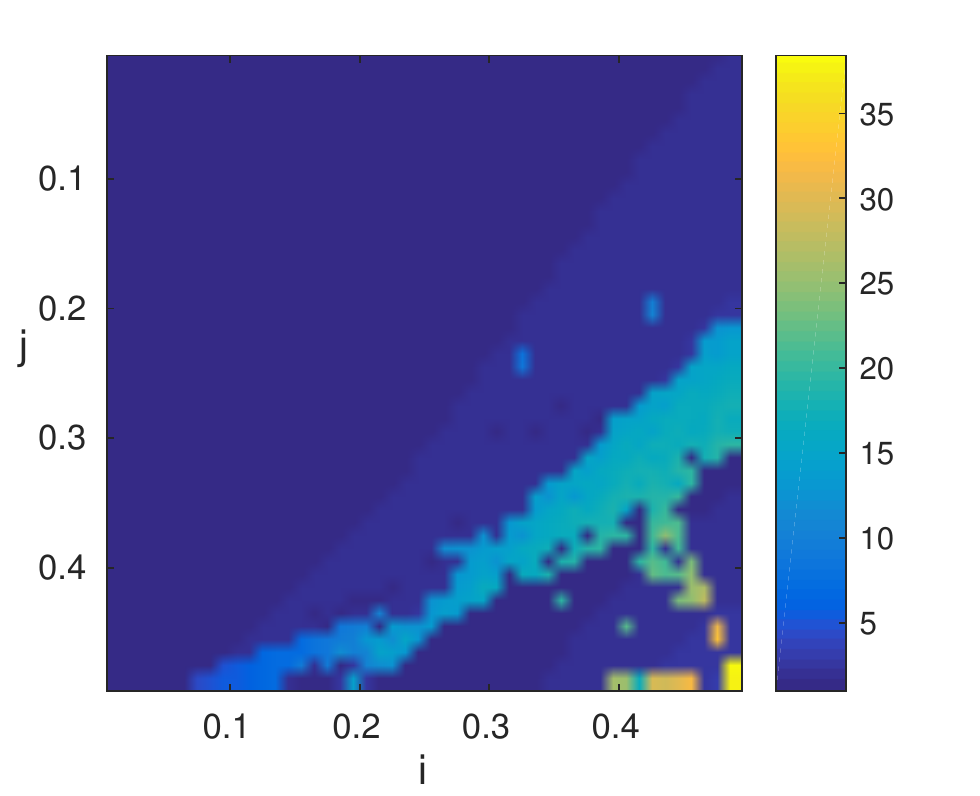}}
\caption{Illustrating the out of plane scaling $\beta$ associated with the microstructure of the given ellipse with array spacing given by figure \ref{fig:Avalue} to attain the $\alpha$ scaling in figure \ref{fig:alphas}}\label{fig:oops}
\end{figure}

\begin{figure}
\centerline{\includegraphics[width=.5\textwidth]{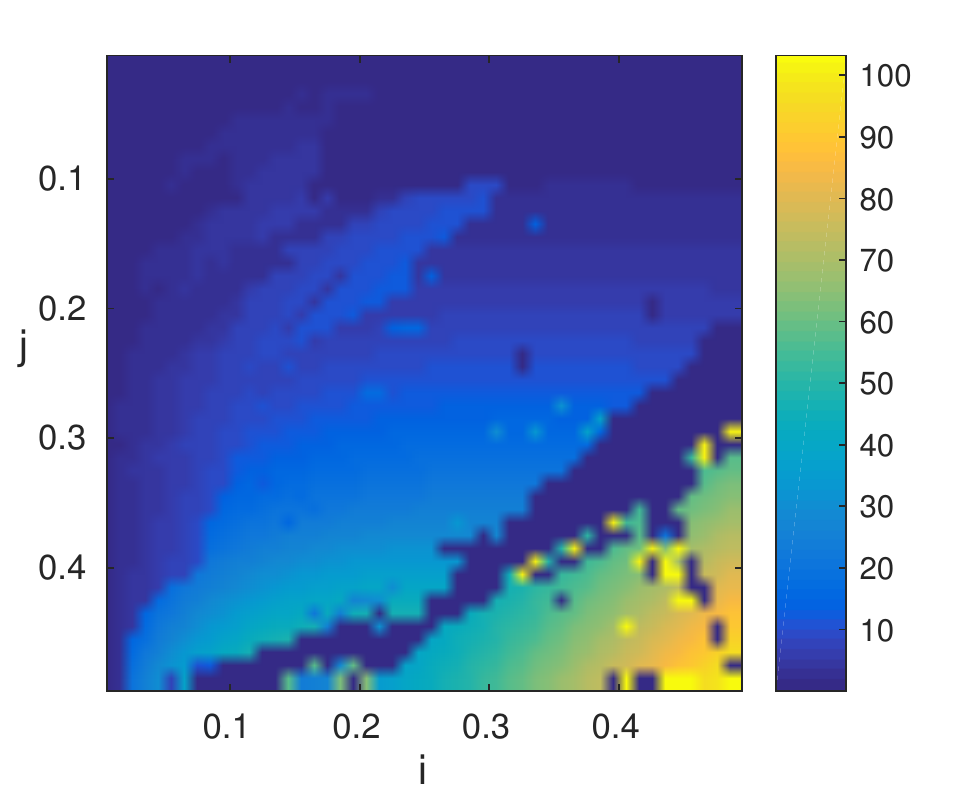}}
\caption{Plot of the percentage error between a given microstructure and the transformation that it is closest to.}\label{fig:pes}
\end{figure}

\begin{figure}
\centerline{\includegraphics[width=.5\textwidth]{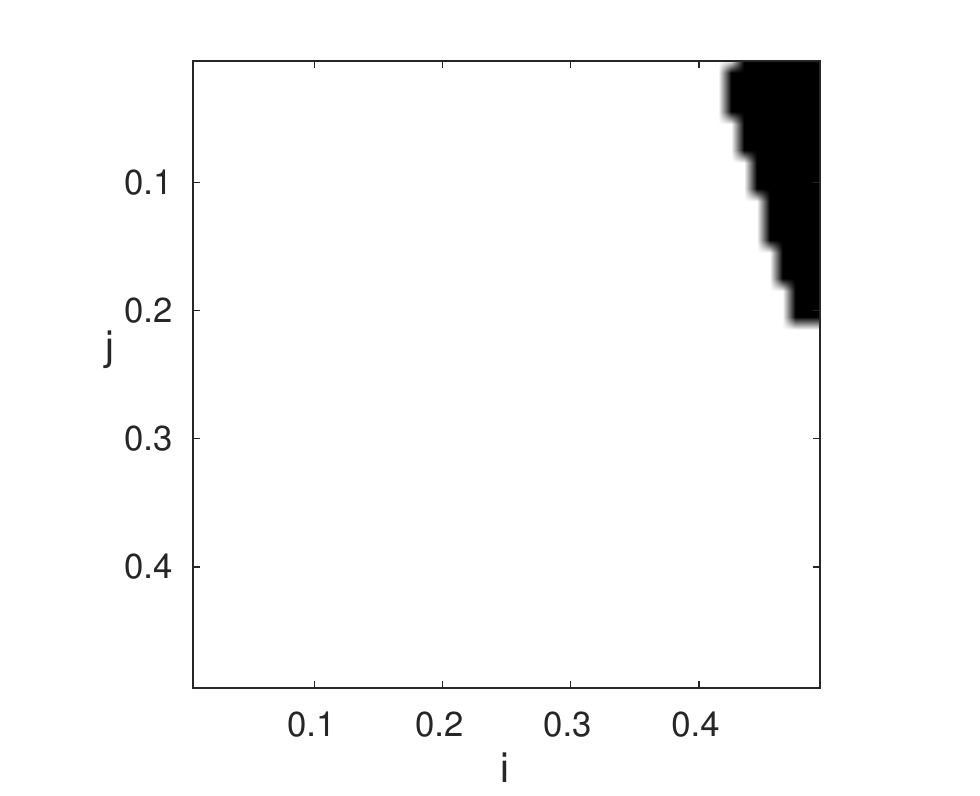}}
\caption{Plot of the region containing desirable ellipses (in black), which are defined by determining whether the microstructure less than 10\% away from their closest transformation. Their closest transformation also has an $\alpha<0.7$ and $\beta<3$.}\label{fig:dr}
\end{figure}

This optimisation scheme is easily automated and indeed fast to run; we can produce plots showing the various parameters found by the scheme in order to display the sensitivity of the transformation that can be achieved when a small change in shape occurs. Each pixel in the plots of Figs \ref{fig:alphas}-\ref{fig:dr} represents a distinct choice of ellipse. The relative semi-axes are specified via \eqref{axi}-\eqref{ayj} with the $i$ and $j$ values specified along the $x$ and $y$ coordinates in the figures. These images specify the ellipse required for each scaling and how sensitive such a microstructure is in achieving this scaling. We show the optimisation results for the transformation properties, $\alpha$ in Fig.\ \ref{fig:alphas} and $\beta$ in Fig.\ \ref{fig:oops}, the necessary relative spacing $A$ is shown in Fig.\ \ref{fig:Avalue}. The remaining plots of this nature show the percentage error between the required and effective properties, Fig.\ \ref{fig:pes}, and a desirable region in Fig.\ \ref{fig:dr}. This desirable region highlights all microstructure that come within 10\% of a transformation where $\alpha<0.7$ and $\beta<3$. These are desirable since the percentage error is low, that is the effect of the microstructure should be as predicted, $\alpha$ is small, and therefore the space elongation is pronounced and $\beta$ is sufficiently close to unity to give a good impedance match. We can see that this desirable region highlights a tight cluster in the upper right hand corner.

Once a region of desirable ellipses has been found mathematically we then use this as a first step to inform an optimisation routine in Comsol Multiphysics. This latter simulation accounts for viscous and thermal losses in the microstructure and its full three-dimensional geometry.

\subsubsection*{Microstructure size and viscous losses}

\begin{figure}[htb]
\centerline{\includegraphics[width=.5\textwidth]{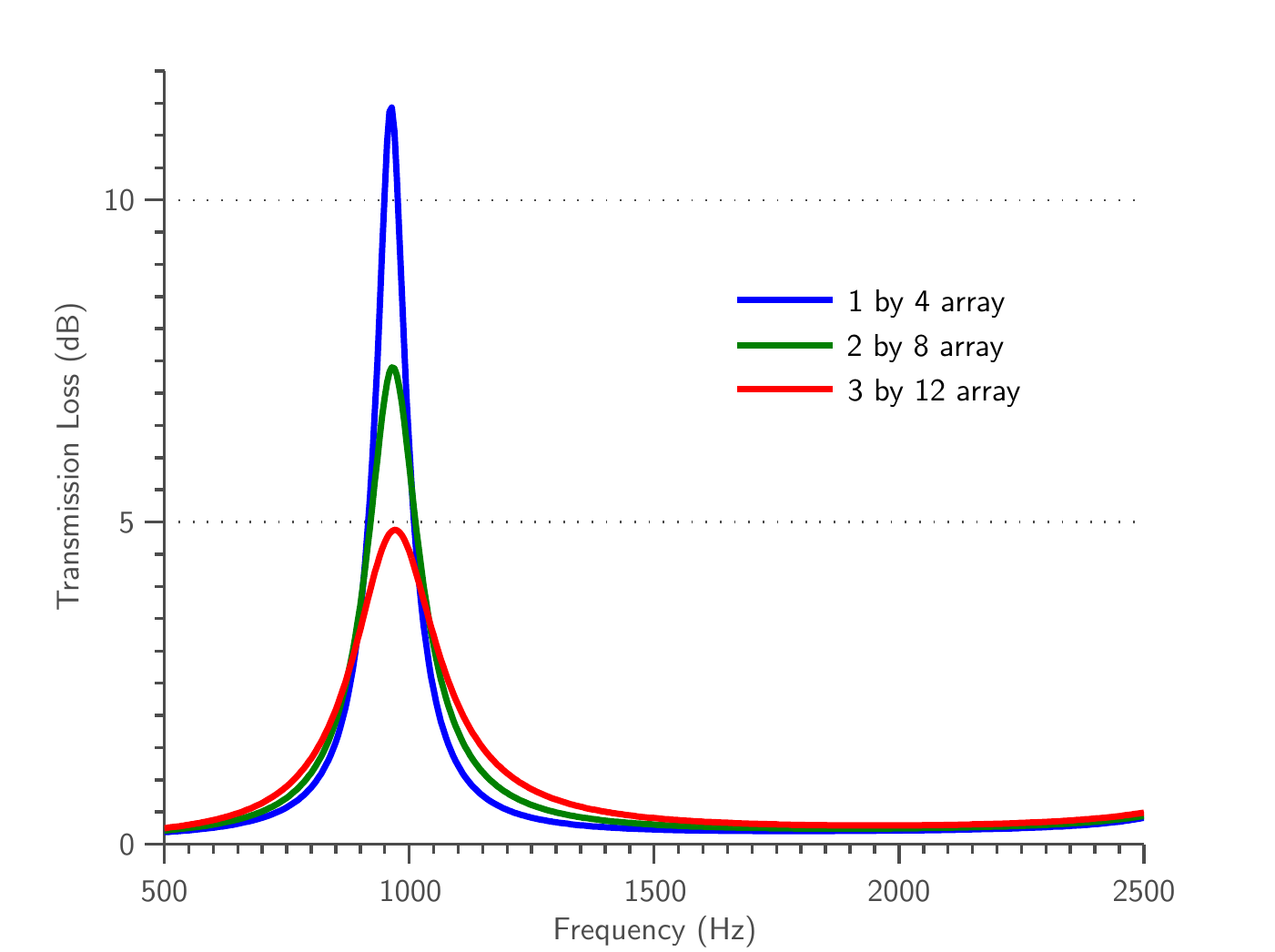}}
	\caption{The simulated transmission loss across a side branch containing a $1\times 4$ array, a $2\times 8$ array and a $3\times 12$ array of elliptical microstructure, when viscous and thermal losses are considered in narrow air gaps.}\label{fig:1v3ellipse}
\end{figure}

In order for the homogenisation result given above to hold, the microstructure must be sufficiently  smaller than the impinging wavelength. Hence the method is applicable to any wavelength as long as one can fabricate a structure that is sufficiently small. However if we construct our ellipses too small then this will increase the surface area of the microstructure and promote viscous and thermal losses in gaps generated by the presence of the ellipses. We have seen in the main body of the paper that viscous losses are undesirable in the QWR application considered here and so must be minimised. We therefore seek an optimal size for the microstructure, where the homogenisation theory is as accurate as possible for the broadest range of frequencies, whilst viscous and thermal losses are not too large. We have implemented numerical simulations of arrays of various sizes of ellipse in the same duct in order to demonstrate this effect; we again use Comsol Multiphysics but specifically the `narrow region' acoustics model. This model takes into account both viscous and thermal losses in the small air gaps between microstructure. Results shown in Fig.\ \ref{fig:1v3ellipse} compare a $1\times 4$,  $2\times 8$ and a $3\times 12$ array of ellipses in a side branch of the same length. We see that we can expect a greater transmission loss for the resonator with larger microstructure, i.e. less viscous and thermal damping occurs here as anticipated. All resonators have a comparable resonant frequency despite the change in microstructure size as predicted. Notice that the transmission loss shown in simulations is greater than that achieved experimentally; these losses are indicative of the true experimental loss since all values will reduce proportionally in reality, even in the case of a resonator without microstructure.

\end{document}